\documentclass[12pt,preprint]{aastex62}
\usepackage{graphicx,natbib,amsmath}
\usepackage{epstopdf}

\newcommand{\lapprox} {\, \lower3pt\hbox{$\sim$}\llap{\raise2pt\hbox{$<$}}\,}
\newcommand{\gapprox} {\, \lower3pt\hbox{$\sim$}\llap{\raise2pt\hbox{$>$}}\,}

\begin{document}

\title{A Fokker-Planck Framework for Studying the Diffusion of Radio Burst Waves in the Solar Corona}

\author{N. H. Bian}
\affil{Department of Physics \& Astronomy, Western Kentucky University, Bowling Green, KY 42101, USA}

\author[0000-0001-8720-0723]{A. G. Emslie}
\affil{Department of Physics \& Astronomy, Western Kentucky University, Bowling Green, KY 42101, USA}

\author[0000-0002-8078-0902]{E. P. Kontar}
\affil{School of Physics \& Astronomy, University of Glasgow, Glasgow G12 8QQ, Scotland, UK}

\begin{abstract}

Electromagnetic wave scattering off density inhomogeneities in the solar corona is an important process which determines both the apparent source size and the time profile of radio bursts observed at 1 AU. Here we model the scattering process using a Fokker-Planck equation and apply this formalism to several regimes of interest. In the first regime the density fluctuations are considered quasi-static and diffusion in wavevector space is dominated by angular diffusion on the surface of a constant energy sphere. In the small-angle (``pencil beam'') approximation, this diffusion further occurs over a small solid angle in wavevector space. The second regime corresponds to a much later time, by which scattering has rendered the photon distribution near-isotropic resulting in a spatial diffusion of the radiation. The third regime involves time-dependent fluctuations and, therefore, Fermi acceleration of photons. Combined, these results provide a comprehensive theoretical framework within which to understand several important features of propagation of radio burst waves in the solar corona: emitted photons are accelerated in a relatively small inner region and then diffuse outwards to larger distances. En route, angular diffusion results both in source sizes which are substantially larger than the intrinsic source, and in observed intensity-versus-time profiles that are asymmetric, with a sharp rise and an exponential decay.  Both of these features are consistent with observations of solar radio bursts.

\end{abstract}

\keywords{Sun: activity -- Sun: corona -- Sun: flares -- Sun: radio radiation}

\section{Introduction}

After the discovery of the interplanetary scintillation phenomenon \citep{1964Natur.203.1214H}, along with its potential for probing the solar wind \citep{1978SSRv...21..411C}, it became undisputably clear that radiative transfer phenomena were crucial to the interpretation of solar radio emission. Transport processes are essential to resolve conundrums such as the excess size and low brightness temperature of quiet-Sun radio emission
\citep{1971A&A....12..435A, 1974SoPh...36..375R, 1977A&A....58..287H, 1992SoPh..140...19T, 1994SoPh..149...31T, 2008ApJ...676.1338T} and they appear to be substantially more important than the properties of the intrinsic source in determining the spatio-temporal characteristics of solar radio bursts \citep{1971A&A....10..362S, 1974SoPh...34..181R, 2007ApJ...671..894T, 1985A&A...150..205S, 2017NatCo...8.1515K}.

Type III bursts are produced when fast ($\sim$$c/3$) electrons accelerated during solar flares travel outward along open magnetic field lines into interplanetary space. The mechanism of emission \citep[e.g.,][]{2017RvMPP...1....5M} involves collective processes in the plasma, specifically the generation of Langmuir waves at the plasma frequency $f_{pe} \simeq 9 \, [n_{e} $(cm$^{-3}$)]$^{1/2}$~kHz and their conversion into electromagnetic waves with frequencies close to the fundamental ($f_{pe}$) or the harmonic ($2f_{pe}$) of the plasma frequency. Since the beam-generated Langmuir waves have frequencies comparable to the local plasma frequency, the properties of the escaping electromagnetic radiation are very sensitive to fluctuations in the local (frequency-dependent) plasma refractive index. Scattering of fundamental mode radiation is expected to be particularly strong in the vicinity of the intrinsic source, resulting in considerable spatial and temporal dispersion of the emitted signal. The importance of scattering was confirmed by the recent high-resolution observations of \cite{2017NatCo...8.1515K}, who showed, using LOFAR imaging spectroscopy of a Type III-b radio burst, that the size of the scattered image was orders of magnitude larger than the intrinsic source size inferred from rapid time variations within the burst striae.

Time profiles of type III bursts have been studied by various authors \citep{1972A&A....19..343A,1973SoPh...29..197A, 1973SoPh...31..501E,1975SoPh...45..459B,1977A&A....56..251P,2018arXiv180201507R}. They are characterized by an asymmetric profile, with a sharp rise and a slower, typically exponential, decay; the asymmetry becomes larger as the exponential decay time becomes larger. Such temporal characteristics share many similarities with those of radio pulsars observed with, e.g., LOFAR \citep{2017MNRAS.470.2659G}. Since angular scattering of photons relative to the line of sight leads to different travel distances \citep{1972MNRAS.157...55W} and hence arrival times at the observer, this suggests that a common scattering process can explain both the angular size and the time profile of solar radio bursts.

The statistical basis for scattering of light rays by random plasma inhomogeneities was first presented by \cite{1952MNRAS.112..475C} and subsequently generalized by \cite{1968AJ.....73..972H}. These theories form the basis of the numerical ray tracing techniques which have been applied to the propagation and scattering of radio waves emitted in the corona and in the interplanetary space \citep{1965BAN....18..111F, 1971A&A....10..362S,1974SoPh...34..181R, 2007ApJ...671..894T,2018ApJ...857...82K}. Another approach, based on the Hamilton equations for photons, was adopted by \cite{1999A&A...351.1165A},  where a Fokker-Planck equation with an energy-conserving diffusion operator was derived and studied. This approach, an alternative to the ray-tracing techniques cited above, provides a powerful tool for both analytical and numerical computational purposes. In this work we therefore extend the analytical description of radio wave propagation in terms of a Fokker-Planck equation that exploits the similarities between photon scattering off density inhomogeneities and particle scattering in magnetized plasmas, a phenomenon which plays a significant role also in the acceleration and transport of energetic electrons during solar flares \citep[e.g.,][]{2011A&A...535A..18B, 2012ApJ...754..103B, 2013PhRvL.110o1101B, 2014ApJ...796..142B, 2014ApJ...780..176K, 2014ApJ...787...86J, 2015ApJ...809...35K,2018ApJ...862..158E}.

The outline of the paper is as follows. In Section~\ref{quasi-linear} we use quasi-linear theory \citep{1963JNuE....5..169V} to derive a general expression for the diffusion tensor in wave-vector space, in which both spatial and temporal variations of the density fluctuations are taken into account. In Section~\ref{angular} we consider the limit where the density fluctuations are quasi-static: scattering is then elastic and the scattering operator appearing in the Fokker-Planck equation reduces to the Lorentz form, in which diffusion in wave-vector space occurs at a constant energy. In Section~\ref{analysis} we analyze the effect of angular scattering on the spatial transport of electromagnetic waves. The Fokker-Planck equation is studied in the elastic regime and in the small angle approximation; the perpendicular and parallel parts of this Fokker-Planck equation are solved separately. The perpendicular part describes the effect of scattering on the spatial spread of photons in the direction perpendicular to the initial direction of propagation.  The results are shown to correspond to Fermi's pencil beam equation \citep{PhysRev.74.1534}, which involves the Kolmogorov operator \citep{Calin2009}; its (Gaussian) solution describes the joint evolution of the photon distribution both in perpendicular wave-number space and in perpendicular space, and from the latter the angular diffusion coefficient in space is readily recovered. The parallel part describes the effect of scattering on the spatial distribution of photons in the direction parallel to the initial direction of propagation; this equation has also a relatively simple form involving the Hermite operator.  Its solution, which is \emph{not} Gaussian, describes the joint evolution of the photon distribution both in perpendicular wave-number space and in parallel space; from the latter the pulse time profile in a scattering medium can be computed.

In Section~\ref{diffusive-analysis} we consider the diffusive regime of spatial transport which results from the cumulative effect of small-angle scattering over many scattering times. We discuss, in this diffusive approximation, both the angular size of the scattered image and the pulse time profile as seen by a distant observer. The pulse time profile is now obtained from the distribution of the first exit times of photons from the scattering region \citep[see][]{2007gfpp.book.....R}. The characteristics of the pulse time profile are, rather surprisingly, very similar to those obtained in the small-angle regime; however, the corresponding exponential decay time is now the spatial diffusion time over the scattering region.

In Section~\ref{fermi-acceleration} we consider temporal variations in density fluctuations located near the intrinsic source of emission; the group velocity of photons here is sufficiently small that even slow temporal variations in the plasma density can cause Fermi-type acceleration \citep{1958PhRv..111.1206P} of the radio photons. In Section~\ref{discussion} we assemble all the results above into a comprehensive picture of photon propagation in solar radio bursts, involving both an inner ``acceleration'' region and an outer ``propagation'' region.

\section{Quasilinear wave diffusion}\label{quasi-linear}

In this section we explore the photon scattering process using a theory originally developed to describe wave-particle interactions \citep{1963JNuE....5..169V}, non-linear wave interactions and wave scattering \citep{1967PlPh....9..719V} in plasmas; see also the monographs by \citet{1969npt..book.....S}, \cite{1995lnlp.book.....T}, and \cite{1980panp.book.....M}. We closely follow the presentation given by \cite{1967PlPh....9..719V} and show how, starting from the wave Hamiltonian, quasilinear theory can be used to describe wave scattering in a fluctuating medium \citep{1961wptm.book.....T,1978wpsr.book.....I}. The methodology is then applied to scattering of the escaping electromagnetic radiation in a manner similar to the scattering of Langmuir waves \citep{2012ApJ...761..176R,2014JGRA..119.4239B}, showing that {\it such an approach serves as a unifying framework for modeling both particle, Langmuir wave and radio wave propagation.} We then use this unified approach to derive a general expression for the diffusion tensor in wave-vector space which accounts for both spatial and temporal variations of the turbulent density fluctuations.

\subsection{Hamilton equations of motion and the wave-kinetic equation}

The dispersion relation for electromagnetic waves in a plasma is

\begin{equation}\label{basic-dispersion-relation}
\omega^{2}(\mathbf{k},\mathbf{r},t)=k^{2}c^{2} + \omega_{pe}^{2}(\mathbf{r},t) \,\,\, ,
\end{equation}
where $\omega$ (s$^{-1}$) is the wave angular frequency and ${\mathbf k}$ (cm$^{-1}$) the wavevector, $\omega_{pe}(\mathbf{r},t) = \sqrt{4\pi n(\mathbf{r},t) \, e^{2}/m_{e}} \, $ (s$^{-1}$) is the plasma frequency, considered to vary in both space and time due to variations in the density $n(\mathbf{r},t)$ (cm$^{-3}$), and $e$ (esu) and $m$ (g) are the electronic charge and mass, respectively. The group velocity $\mathbf{v}_{g}$ (cm~s$^{-1}$) of the electromagnetic radiation is

\begin{equation}\label{group-velocity}
\mathbf{v}_{g} \equiv \frac{\partial \omega}{\partial \mathbf{k}}=\frac{c^{2}}{\omega} \, \mathbf{k}= \frac{c^{2}}{\sqrt{k^{2}c^{2}+\omega_{pe}^{2}}} \, \mathbf{k}
\end{equation}
and the plasma refractive index is given by

\begin{equation}\label{refractive}
\mu_{r} \equiv \frac{c}{v_{p}}=\frac{v_{g}}{c}=\left ( 1-\frac{\omega_{pe}^{2}} {\omega^{2}}\right )^{1/2} \,\,\, ,
\end{equation}
where $v_{p}=\omega/k>c$ is the phase speed.
Geometrical optics studies of scattering of radio waves \citep[e.g.,][]{1952MNRAS.112..475C, 1968AJ.....73..972H,1970JGR....75.3715H,1998ApJ...506..456C}
have been based on a perturbation of the eikonal equation

\begin{equation}\label{rayban}
\frac{d}{d s} \left ( \mu_{r} \, \frac{d\mathbf{r}}{d s} \right ) = \frac{\partial \mu_{r}}{\partial \mathbf{r}}\,\,\, ,
\end{equation}
which, in the small-angle approximation, can be reduced to two equations:

\begin{equation}\label{rayban1}
\frac{d \Psi_{1}}{ds}=\frac{1}{\overline{\mu}_{r}}\frac{d\delta \mu_{r}}{dx}  \,\,\, , \qquad \frac{d \Psi_{2}}{ds}=\frac{1}{\overline{\mu}_{r}}\frac{d\delta \mu_{r}}{dy} \,\,\, ,
\end{equation}
for the angles $\Psi_{1}=dx/ds$ and $\Psi_{2}=dy/ds$ characterizing the directions by which the ray deviates from its unperturbed direction due to small perturbations in the refractive index $\delta \mu_{r}$ around its mean value $\overline{\mu}_{r}$. Such a small-angle analysis can be used to obtain an analytical expression for the mean square angular deviation of the ray, $\langle \Psi^{2} \rangle=\langle \Psi_{1}^{2} \rangle +\langle \Psi_{2}^{2}\rangle =2\langle \Psi_{1}^{2}\rangle $, in a slab of thickness $\Delta s$:

\begin{equation}\label{smr}
\langle \Psi^{2}\rangle = D_{\Psi} \, \Delta s \,\,\, ,
\end{equation}
where $D_{\Psi}$ (cm$^{-1}$) is the angular diffusion coefficient. In the seminal work by \cite{1952MNRAS.112..475C}, only the case $\overline{\mu}_{r}\simeq 1$, i.e., $\omega \gg \overline{\omega}_{pe}$, corresponding to the weak scattering regime (see below), is considered. Numerical ray-tracing studies of radio-waves \citep[e.g.,][]{2007ApJ...671..894T,2018ApJ...857...82K} have been based on a set of first-order ordinary differential equations \citep{1963JATP...25..397H}:

\begin{equation}\label{h1}
\frac{d\mathbf{r}}{d\tau}=\mathbf{u} \,\,\, ,
\end{equation}

\begin{equation}\label{h2}
\frac{d\mathbf{u}}{d\tau}=\frac{1}{2} \,\frac{\partial \overline{\mu}_{r}^{2}}{\partial \mathbf{r}} \,\,\, ,
\end{equation}
for the position $\mathbf{r}$ and direction $\mathbf{u}$ of the ray as a function of $\tau$, where $d\tau=ds/\overline{\mu}_{r}$. These equations are easily obtained from Equation~(\ref{rayban}) and they describe the ray path in a medium with smoothly varying index of refraction $\overline{\mu}_{r}$. The role of scattering, resulting from the fluctuations $\delta \mu_{r}$, is implemented by adding random gaussian perturbations to the ray direction $\mathbf{u}$ at each step $\Delta \tau$ in the numerical integration of Equations (\ref{h1}) and~(\ref{h2}), with the standard deviation of the perturbations taken from the small-angle result (\ref{smr}).

Here, following \cite{1999A&A...351.1165A}, we instead base our study on the Hamilton equations describing the joint evolution of the wave-vector $\mathbf{k}$ and the position $\mathbf{r}$ of radio photons as a function of time $t$. In the WKB framework, the wave Hamiltonian is simply the wave frequency $\omega = E/\hbar$, with $\mathbf{p}=\hbar \, \mathbf{k}$ the wave momentum, and therefore the photon trajectories are governed by the Hamilton equations

\begin{equation}\label{rdot}
\dot{\mathbf{r}}=\frac{\partial \omega}{\partial \mathbf{k}} \equiv \mathbf{v}_g
\end{equation}
and

\begin{equation}\label{kdot}
\dot{\mathbf{k}} = - \frac{\partial \omega}{\partial \mathbf{r}} \,\,\, .
\end{equation}

We represent the electron density $n(\mathbf{r},t)$ as the sum of a large-scale, non-uniform but time-independent, background plus small-scale fluctuations that depend on both position and time:

\begin{equation}\label{n-components}
n(\mathbf{r},t) = \overline{n}(\mathbf{r})+\widetilde{n}(\mathbf{r},t) \,\,\, .
\end{equation}
Then $\omega^{2}_{pe} = \overline{\omega}^{2}_{pe} (1+\widetilde{n}/\overline{n})$, and, in the limit $\widetilde{n} \ll \overline{n}$, the wave Hamiltonian can be split into two parts:

\begin{equation}\label{hamiltonian-split}
\omega=\overline{\omega} + \widetilde{\omega} \simeq \sqrt{k^{2}c^{2}+\overline{\omega}_{pe}^2}
 + \frac{1}{2} \, \frac{\overline{\omega}_{pe}^{2}}  {\overline{\omega}} \, \frac{\widetilde{n}}{\overline{n}} \,\,\, .
\end{equation}
The evolution of the phase-space distribution $N({\mathbf r},\mathbf{k},t)$ of photons is then governed by a wave-kinetic equation written in the form

\begin{equation}\label{liouville}
\frac{\partial N}{\partial t}
+\mathbf{v}_{g} \cdot\frac{\partial N}{\partial \mathbf{r}}
+(\overline{\mathbf{f}} + \widetilde{\mathbf{f}}) \cdot \frac{\partial N}{\partial \mathbf{k}}=0 \,\,\, .
\end{equation}
Here the ``refractive force'' $\mathbf{f} \equiv \dot{\mathbf{p}}/\hbar$ (cm$^{-1}$~s$^{-1}$) associated with change of wave momentum has two components, one a relatively smooth large-scale force

\begin{equation}\label{overline-f-exp}
\overline{\mathbf{f}}({\mathbf r}) = - \frac{\partial \overline{\omega}}{\partial \mathbf{r}} =
- \frac{1}{2 \, \overline{\omega}} \, \frac{\partial \overline{\omega}_{pe}^{2}}{\partial \mathbf{r}}
\end{equation}
that leads to refraction, and a small-scale fluctuating component

\begin{equation}\label{tilde-f-exp}
\widetilde{\mathbf{f}}({\mathbf r}, t) = - \frac{\partial \widetilde{\omega}}{\partial \mathbf{r}} = -\frac{1}{2} \, \frac{\overline{\omega}_{pe}^{2}}{\overline{\omega}} \, \frac{\partial }{\partial \mathbf{r}} \left ( \frac{\widetilde{n}({\mathbf r}, t)}{\overline{n} ({\mathbf r})} \right )
\end{equation}
that results in scattering. It should be kept in mind, however, that both the refractive and scattering terms have a common physical origin.

\subsection{The Fokker-Planck equation and quasilinear diffusion in wave-vector space}

Under fairly general conditions \citep[e.g.,][]{1979PhR....52..263C, 1983rsm..book.....L, 1990A&A...232..215Z}, the scattering term in the wave-kinetic equation produces a diffusion of photons in momentum space, a phenomenon known as stochastic acceleration \citep{1958PhRv..111.1206P, 1966PhRv..141..186S, 1997JGR...10214631M, 2012ApJ...754..103B}.  Such an evolution of the phase-space distribution can be described by a Fokker-Planck equation

\begin{equation}\label{fp-equation}
\frac{\partial N}{\partial t} + \mathbf{v}_{g} \cdot \frac{\partial N}{\partial \mathbf{r}} + \overline{\mathbf{f}}\cdot \frac{\partial N}{\partial \mathbf{k}}= \frac{\partial}{\partial k_{i}} \left [ D_{ij}(\mathbf{k}) \, \frac{\partial N}{\partial k{j}} \right ] \,\,\, ,
\end{equation}
where $D_{ij}(\mathbf{k})$ is the diffusion tensor in wave-vector space. Notice that the phase-space distribution $N(\mathbf{r},\mathbf{k},t)$ here has a somewhat different meaning in the Fokker-Planck equation~(\ref{fp-equation}) than in the original wave-kinetic equation~(\ref{liouville}): in Equation~(\ref{liouville}), $N$ represents the exact fine-grained phase space distribution, while in the Fokker-Planck equation~(\ref{fp-equation}) $N$ represents a coarse-grained phase space distribution. We shall, however, use the same notation for both quantities, since no confusion will arise.

We now proceed to obtain, via the quasilinear approximation, the expression for the tensor describing diffusion of photons in wave-vector space. From the relation $\Delta k_{i}(t)=\int _{0}^{t}ds \, \widetilde{f}_{i}(s)$, with $\widetilde{f}_{i}(s)=\widetilde{f}_{i}(\mathbf{r}(s),s)$, we find \citep[cf. the derivation of Equation~(1.4) in][]{Bleher1992} the expression

\begin{equation}\label{deltak-deltak}
\langle \Delta k_{i} \, \Delta k_{j} \rangle (t) = 2t\int_{0}^{t} ds \, \langle \widetilde{f}_{i}(0)\, \widetilde{f}_{j}(s) \rangle - 
2\int_{0}^{t} s \, \langle \widetilde{f}_{i}(0)\, \widetilde{f}_{j}(s) \rangle \,  ds \,\,\, ,
\end{equation}
where $\langle \cdots \rangle $ denotes the ensemble average.  Assuming that the Lagrangian auto-correlation of the fluctuating force decays sufficiently rapidly, this leads to the usual Taylor-Green-Kubo formula \citep{taylor1921gi}\footnote{the original work by G.I. Taylor is aimed at describing the spatial dispersion of a passive tracer in a turbulent fluid; the dispersion, here of photons, occurs in phase-space.} in the limit $t\rightarrow\infty$. The diffusion tensor (cm$^{-2}$~s$^{-1}$) in wave-vector space is then formally given by the expression

\begin{equation}\label{dif-def}
D_{ij}\equiv \lim _{t\rightarrow \infty} \frac{ \langle \Delta k_{i}\Delta k_{j} \rangle (t)}{2t}= \int_{0}^{\infty} dt \, \langle \widetilde{f}_{i}(0,0) \, \widetilde{f}_{j}(\mathbf{r(t)},t) \rangle \,\,\, ,
\end{equation}
involving the Lagrangian correlations of the fluctuating force, taken along photon trajectories. The density fluctuations are characterized by their {\it Eulerian} auto-correlation function (cm$^{-6}$)

\begin{equation}\label{eulerian-auto-correlation}
C_{\widetilde{n}}(\mathbf{r},t)=\langle \, \widetilde{n}(0,0) \, \widetilde{n}(\mathbf{r},t) \, \rangle \,\,\,
\end{equation}
and the Eulerian statistical properties of the small-scale fluctuating part $\widetilde{\mathbf{f}}$ of the refractive force are characterized by the correlation tensor (cm$^{-2}$~s$^{-2}$):

\begin{equation}\label{f-formal}
C_{ij}(\mathbf{r},t)=\langle \, \widetilde{f}_{i}(0,0) \, \widetilde{f}_{j}(\mathbf{r},t) \, \rangle  \,\,\, .
\end{equation}
Since the refractive force $\widetilde {\bf f}$ is proportional to the fluctuating density gradient (Equation~(\ref{tilde-f-exp})), this correlation tensor is related to $C_{\widetilde{n}}$ via

\begin{equation}\label{cij-def}
C_{ij} ({\mathbf r}, t)= \frac{\overline{\omega}_{pe}^{4}}{4 \, \overline{\omega}^{2}} \, \frac{1}{\overline{n}^{2}} \, \frac{\partial^{2} C_{\widetilde{n}} ({\mathbf r}, t)}{\partial r_{i} \, \partial {r_{j}}} \,\,\, .
\end{equation}
From Equation~(\ref{dif-def}), the diffusion tensor in wave-vector space is therefore re-written in the form

\begin{equation}\label{dif-def-bis}
D_{ij}=\frac{\overline{\omega}_{pe}^{4}}{4 \, \overline{\omega}^{2}} \, \frac{1}{\overline{n}^{2}} \, \int_{0}^{\infty} dt \, \left . \frac{\partial^{2} C_{\widetilde{n}} ({\mathbf {r}}, t)}{\partial r_{i} \, \partial {r_{j}}} \right \vert_{\mathbf{r} = \mathbf{r}(t)} \,\,\, .
\end{equation}
This can be more conveniently expressed in terms of spectral quantities. Let us therefore decompose the fluctuating fields into Fourier modes, according to the convention

\begin{equation}\label{fourier-decomp}
\widetilde{n}(\mathbf{r},t)=\int d\mathbf{q} \, d\Omega \, \widetilde{n}(\mathbf{q},\Omega) \, e^{i(\mathbf{q}.\mathbf{r}-\Omega t)} \,\,\, .
\end{equation}
The two-point correlation function can then be written

\begin{equation}\label{cfn}
\langle \widetilde{n}(\mathbf{r},t) \, \widetilde{n}(\mathbf{r}',t') \rangle = \int d\mathbf{q} \, d\mathbf{q}' \, d\Omega \, d\Omega' \, \langle \widetilde{n}(\mathbf{q},\Omega) \, \widetilde{n}(\mathbf{q}',\Omega')\rangle \, e^{i(\mathbf{q}.\mathbf{r}+\mathbf{q}'.\mathbf{r}'-\Omega t-\Omega' t')} \,\,\, .
\end{equation}
We assume that the density fluctuations are statistically homogeneous and stationary, meaning that the right-hand side of this expression must depend only on $(\mathbf{r}-\mathbf{r}')$ and $(t-t')$.  This is possible only if $\mathbf{q} = - \mathbf{q}'$ and $\Omega=-\Omega'$, so that the average value $\langle n(\mathbf{q},\Omega) \, n(\mathbf{q}',\Omega') \rangle$ is proportional to $\delta (\mathbf{q}+\mathbf{q'})$ and $\delta(\Omega+\Omega')$:

\begin{equation}\label{spn}
\langle \widetilde{n}(\mathbf{q},\Omega) \, \widetilde{n}(\mathbf{q}',\Omega')\rangle = \tilde{n}|^{2}_{\mathbf{q},\Omega} \, \delta (\mathbf{q}+\mathbf{q'}) \, \delta(\Omega+\Omega') \,\,\, ,
\end{equation}
which serves as the definition of the spectrum $\tilde{n}|^{2}_{\mathbf{q},\Omega}$ of density fluctuations. Indeed, inserting Equation~(\ref{spn}) into Equation~(\ref{cfn}), we readily obtain the Wiener-Khinchin theorem

\begin{equation}\label{wiener-khinchin}
C_{\widetilde{n}}(\mathbf{r},t)=\int d\mathbf{q} \, d\Omega \,  \tilde{n}|^{2}_{\mathbf{q},\Omega} \, e^{i(\mathbf{q}.\mathbf{r}-\Omega t)} \,\,\, ,
\end{equation}
and inserting this in Equation~(\ref{dif-def-bis}), we obtain the expression

\begin{equation}\label{dij-exp-1}
D_{ij}(\mathbf{k}) = \frac{ \overline{\omega}_{pe}^{4}}{4 \, \overline{\omega}^{2}} \, \int d\Omega \, d\mathbf{q} \,
q_{i} q_{j} \, \left . \frac{\tilde{n}}{\overline{n}} \right \vert ^{2}_{\mathbf{q},\Omega} \, \int_{0}^{\infty} dt \, e^{i(\mathbf{q}.\mathbf{r}(t)-\Omega t)}
\end{equation}
for the diffusion tensor in wavevector space.

The quasilinear approximation consists of setting $\mathbf{r}(t) = \mathbf{v}_{g} \, t$ when evaluating\footnote{This is analogous to the Born approximation in scattering theory, where, in computing the change in momentum due to a given force field, one takes the momentum to be a constant in calculating the acceleration.} the time integral in Equation~(\ref{dij-exp-1}):

\begin{equation}\label{quasilinear-time-integral}
\int_{0}^{\infty} dt \, e^{i(\mathbf{q}.\mathbf{r}(t)-\Omega t)} = \int_{0}^{\infty} dt \, e^{i(\mathbf{q}.\mathbf{v}_{g}-\Omega) t} = \pi \, \delta(\Omega-\mathbf{q}.\mathbf{v}_{g}(\mathbf{k})) \,\,\, ,
\end{equation}
so that

\begin{equation}\label{dij-exp}
D_{ij}(\mathbf{k}) = \frac{\pi \, \overline{\omega}_{pe}^{4}}{4 \, \overline{\omega}^{2}} \, \int d\Omega \, d\mathbf{q} \,
q_{i} q_{j} \, \left . \frac{\tilde{n}}{\overline{n}} \right \vert ^{2}_{\mathbf{q},\Omega} \, \delta(\Omega-\mathbf{q}.\mathbf{v}_{g}(\mathbf{k})) \,\,\, .
\end{equation}
This is the general expression for the diffusion tensor in wave-vector space in the quasilinear approximation, where both spatial ($\mathbf{q}$) and temporal ($\Omega$) variations of the density fluctuations are taken into account. From this general expression more particular forms may be obtained. For example, if the density fluctuations are produced by a compressive plasma mode with dispersion relation $\Omega(\mathbf{q})$, implying a density spectrum

\begin{equation}\label{density-spectrum}
\tilde{n}|^{2}_{\mathbf{q},\Omega}= \tilde{n}|^{2}_{\mathbf{q}} \, \delta
(\Omega-\Omega(\mathbf{q}))
\end{equation}
that is strongly peaked at the relevant mode frequency, then the quasilinear diffusion tensor takes the form

\begin{equation}\label{dij-expression}
D_{ij}(\mathbf{k}) = \frac{\pi \, \overline{\omega}_{pe}^{4}}{4 \, \overline{\omega}^{2}} \int d\mathbf{q} \,
q_{i} q_{j} \, \left . \frac{\tilde{n}}{\overline{n}} \right \vert ^{2}_{\mathbf{q}} \, \delta(\Omega(\mathbf{q})-\mathbf{q}.\mathbf{v}_{g}(\mathbf{k})) \,\,\, .
\end{equation}
This result can be interpreted in terms of non-linear mode coupling between waves. Consider an electromagnetic wave scattering (i.e., refracting) off turbulent plasma waves.  Energy conservation requires that

\begin{equation}\label{energy-conservation}
\omega(\mathbf{k})+\Omega(\mathbf{q}) = \omega \, (\mathbf{k}+\mathbf{q}) \,\,\, .
\end{equation}
If $\mathbf{q} \ll \mathbf{k}$, i.e., there are no large changes in the electromagnetic wave momentum (either in the direction or the absolute value of the wave-vector $\mathbf{k}$), then a Taylor expansion of Equation~(\ref{energy-conservation}) gives the resonance condition

\begin{equation}\label{res}
\Omega(\mathbf{q})-\mathbf{q}.\mathbf{v}_{g}=0 \,\,\, ,
\end{equation}
which is the argument of the $\delta$-function appearing in the expression~(\ref{dij-expression}) for the diffusion tensor in wave-vector space. Note from Equation~(\ref{energy-conservation}) that refraction off time-dependent fluctuations generally implies a change in frequency.

Overall, then, the evolution of the phase-space distribution $N({\mathbf k},{\mathbf r},t)$ of photons is governed by a Fokker-Planck equation

\begin{equation}\label{fp-equation-f}
\frac{\partial N}{\partial t}
+\frac{c^{2}}{\omega} \, \mathbf{k} \cdot \frac{\partial N}{\partial \mathbf{r}}=\frac{1}{2 \, \overline{\omega}}
\frac{\partial \overline{\omega}_{pe}^{2}}{\partial \mathbf{r}} \cdot \frac{\partial N}{\partial \mathbf{k}}+
\frac{\partial}{\partial k_{i}} \left [ D_{ij}(\mathbf{k}) \, \frac{\partial N}{\partial k{j}} \right ] \,\,\,,
\end{equation}
where the second term on the left-hand side describes spatial streaming, the first term on the right-hand side describes refraction and the last term describes scattering of photons. The right-hand side of this equation therefore accounts for both regular and stochastic change in the radiation wave-vector, the latter being modeled by a diffusion.

In the quasilinear approximation, the wave-vector diffusion tensor $D_{ij}(\mathbf{k})$ is given by Equation~(\ref{dij-exp}) and this tensor can be decomposed into two components, one perpendicular and one parallel to the radiation wave-vector~$\mathbf{k}$, i.e.,

\begin{equation}\label{dperp-dparallel}
D_{ij}(\mathbf{k}) = D_{\perp}(k) \left ( \delta_{ij} - \frac{k_{i}k_{j}}{k^{2}} \right ) + D_{\parallel}(k) \, \frac{k_{i}k_{j}}{k^{2}} \,\,\, .
\end{equation}
In a spherical coordinate system in wave-vector space, the diffusion operator then reads

\begin{equation}\label{diffusion-spherical}
\frac{\partial}{\partial k_{i}} \left [ D_{ij}(\mathbf{k})\frac{\partial N}{\partial k{j}} \right ] = \frac{1}{k^{2}} \, \frac{\partial
}{\partial k} \left [ k^{2}D_{\parallel}\frac{\partial N}{\partial k} \right ] + \frac{D_{\perp}}{k^{2}} \left [ \frac{1}{\sin \theta} \, \frac{\partial }{\partial \theta}
\left ( \sin \theta \, \frac{\partial N}{\partial \theta} \right ) + \frac{1}{\sin^{2} \theta} \,
\frac{\partial^{2}N}{\partial \phi^{2}} \right ]  \,\,\, .
\end{equation}
Scattering by density fluctuations is therefore in general characterized by \emph{two} scalars $D_{\parallel}$ and~$D_{\perp}$ entering the diffusion tensor in wave-vector space: $D_{\parallel}$ is the diffusion coefficient in $k$, the absolute value of the wave-vector ${\bf k}$, while $D_{\perp}$ is the diffusion coefficient in the two angles $\theta$ and $\phi$ characterizing the direction of the wave-vector $\bf {k}$, i.e., the angular diffusion coefficient in wave-vector space.

\subsubsection{Emission and absorption of electromagnetic waves}\label{emission-mechanisms}

As it stands, the Fokker-Plank equation conserves the total number of photons $N_{p} (t) =\int d\mathbf{r} \, d\mathbf{k} \, N({\mathbf k},{\mathbf r},t)$, a property which allows us to interpret the phase-space density $N({\mathbf k},{\mathbf r},t)$, normalized to $N_{p}$, as a probability distribution function. In order to obtain a more complete description of radiative transfer we need to also include terms describing the rates of emission and absorption of photons. The radiative transport equation, accounting for sources and sinks of radiation is

\begin{equation}\label{transport-equation-f}
\frac{\partial N}{\partial t}
+\frac{c^{2}}{\omega} \, \mathbf{k} \cdot \frac{\partial N}{\partial \mathbf{r}}=\frac{1}{2 \, \overline{\omega}}
\frac{\partial \overline{\omega}_{pe}^{2}}{\partial \mathbf{r}} \cdot \frac{\partial N}{\partial \mathbf{k}}+
\frac{\partial}{\partial k_{i}} \left [ D_{ij}(\mathbf{k}) \, \frac{\partial N}{\partial k{j}} \right ] - \nu_{a} \, N + R \,\,\, ,
\end{equation}
where $\nu_{a}$ and $R$ are the electromagnetic wave absorption coefficient, and emission rate, respectively.

There are a couple of difficulties involved in providing a fully consistent description of emission and absorption of solar radio burst radiation. A main one is that the plasma emission mechanism responsible for radio bursts is a collective plasma process: the intrinsic sources of photons are not discrete particles but rather spatially extended Langmuir waves of relatively high amplitude. The rate of emission, encapsulated in the term $R$ in the above transport equation therefore involve non-linear mode coupling phenomena, which are different for fundamental and harmonic radiation. According to \cite{1998SoPh..181..363R} and \cite{2000SoPh..194..345R}, the rate $R^{F}$ of fundamental emission is dominated by the process of electromagnetic decay of the beam-driven Langmuir waves stimulated by sound waves:

\begin{equation}
L + S \rightarrow F \,\,\, ,
\end{equation}
while the rate $R^{H}$ of harmonic emission results from the coalescence process of two Langmuir waves:

\begin{equation}
L + L' \rightarrow H \,\,\, .
\end{equation}
The corresponding emission rates $R^F$ and $R^H$ are thus non-linear functionals of the Langmuir wave number density $N_{L}$ (not to be confused with the electromagnetic wave number density $N \equiv N_{EM}$ in the transport equation (\ref{transport-equation-f})). The presence of background density fluctuations produces an intermittent, clumpy spatial distribution of this source \citep[e.g.,][]{1979ApJ...233..998S,1985SoPh...96..181M,1992SoPh..139..147R,2014JGRA..119.4239B,2017A&A...598A..44R}. \cite{2006PhRvL..96n5005L,2008JGRA..113.6104L,2008JGRA..113.6105L} and \cite{2014A&A...572A.111R} have numerically evolved the complete set of coupled transport equations for the electron distribution (the ultimate reservoir of free energy) and the wave distributions ($N_{L},N_{S}, N_{H}$ and $N_{F}$); \citep[see also][for particle-in-cell simulations]{2014PhPl...21l2104R, 2018arXiv181102392L}. However, due to numerical limitations, the transport of energy in these models is often assumed to occur in highly simplified geometries. By focusing on the transport of electromagnetic wave energy alone, we can relax these geometrical restrictions.

We shall consider the source of emission to be point-wise both spatially and temporally, so that Equation~(\ref{transport-equation-f}) generates the Green function describing the propagation of a monochromatic impulse of electromagnetic energy

\begin{equation}\label{point-source}
R =  \delta (\mathbf{r}-\mathbf{r}_{0}) \, \delta (\mathbf{k}-\mathbf{k}_{0}) \, \delta (t-t_{0}) \,\,\, .
\end{equation}
Once the propagator $G(\mathbf{r},\mathbf{k},t; \mathbf{r}_{0},\mathbf{k}_{0},t_{0})$ corresponding to this point source function has been determined, different spatio-temporal distributions $R_{0}(\mathbf{r}_{0},\mathbf{k}_{0},t_{0})$ of the intrinsic source of radiation can then be incorporated into the model through the relation

\begin{equation}\label{gf}
N(\mathbf{r},\mathbf{k},t) = \int \int \int d\mathbf{r}_{0} \, d\mathbf{k}_{0} \, dt_{0} \, G(\mathbf{r},\mathbf{k},t; \mathbf{r}_{0},\mathbf{k}_{0},t_{0}) \, R_{0}(\mathbf{r}_{0},\mathbf{k}_{0},t_{0}) \,\,\, .
\end{equation}
The wavevector ($\mathbf{k}_{0}$) dependence
of the source distribution $R_{0}(\mathbf{r}_{0},\mathbf{k}_{0},t_{0})$ is related to the specific radiation patterns (e.g., dipole or quadrupole) associated with either fundamental or harmonic radiation. Despite plasma emission being fundamentally a non-linear process, we stress that the transport of radio waves from the source to the observer is still described by the \emph{linear} equation (\ref{transport-equation-f}). This opens the possibility of studying the intrinsic properties of the source by solving the inverse problem corresponding to Equation~(\ref{gf}) to determine the spatial and temporal properties of the intrinsic source $R_{0}$ from the observed properties of the emitted photons. There are, however, no known general analytical solutions to Equation (\ref{transport-equation-f}), and so analytic forms of the propagator kernel $G(\mathbf{r},\mathbf{k},t; \mathbf{r}_{0},\mathbf{k}_{0},t_{0})$, even for a point source. Hence we will below focus on particular regimes of electromagnetic wave transport, e.g., small-angle or diffusive.

Absorption of the electromagnetic waves can take various forms, all of which essentially involve inverse processes to those associated with wave emission. For example, collisional absorption is a small effect when considering the transport of interplanetary bursts but needs to be considered lower in the corona ($\sim$100~MHz) where the density is higher. Collisional absorption is characterized by the plasma collision frequency

\begin{equation}\label{collision-frequency}
\nu_{c} \simeq  \frac{\pi n_{e}^{4} \ln \Lambda}{m_{e}^{2} \, v_{te}^{3}} \,\,\, ,
\end{equation}
which is also the approximate collisional damping rate of Langmuir waves; see \cite{2016PhPl...23f4504T}.
This parameter thus controls the collisional decay of the source emission. The parameter which governs the rate of collisional absorption of photons by the inverse Bremsstrahlung process, denoted by $\nu_{a}$ in Equation (\ref{transport-equation-f}), is related to the collision frequency $\nu_{c}$ via

\begin{equation}\label{collisional-absorption-frequency}
\nu_{a}=\frac{\omega_{pe}^{2}}{\omega^{2}} \, \nu_{c} \,\,\, .
\end{equation}
Although the collisional absorption term in the transport equation (\ref{transport-equation-f}) results in a decrease of the number of photons, it does not produce a spatial dispersion of an emitted electromagnetic pulse. Rather, spatial dispersion is controlled by the photon-number-conserving transport terms in Equation~(\ref{transport-equation-f}).

\section{Angular diffusion in wave-vector space}\label{angular}

We now consider the limiting case where the wave group velocity is large compared to the phase velocity of the density fluctuations, i.e.,

\begin{equation}\label{static-condition}
v_{g}\gg \frac{\Omega}{q} \,\,\, .
\end{equation}
For such a case, the resonance condition (\ref{res}) reduces to

\begin{equation}\label{static-approximation}
\mathbf{q}.\mathbf{v}_{g} \simeq 0 \,\,\, ,
\end{equation}
meaning that in a frame comoving with a photon at velocity ${\bf v}_{g}$, the density fluctuations are effectively static.  We can thus set $\Omega=0$ in Equation~(\ref{energy-conservation}) to obtain

\begin{equation}\label{wavenumber-before-after}
\omega(\mathbf{k}+\mathbf{q})=\omega(\mathbf{k}) \,\,\, ,
\end{equation}
Since the photon frequency depends only on the wavenumber magnitude $k$, both the photon energy $\hbar \omega$ and momentum magnitude $\hbar k$ are conserved during the scattering process, as in a Lorentz gas. For such zero-frequency scattering modes,

\begin{equation}\label{n-tilde-static}
\widetilde{n}|^{2}_{\mathbf{q},\Omega} = \widetilde{n}|^{2}_{\mathbf{q}} \, \delta(\Omega) \,\,\, ,
\end{equation}
and the diffusion tensor reduces to

\begin{equation}\label{dij-fundamental-result}
D_{ij}(\mathbf{k})=\frac{\pi \, \overline{\omega}_{pe}^{4}}{4 \, \overline{\omega}^{2}} \int d\mathbf{q} \,
q_{i} q_{j} \, \left . \frac{\widetilde{n}}{\overline{n}} \right \vert^{2}_{\mathbf{q}} \, \delta(\mathbf{q}.\mathbf{v}_{g}) \,\,\, .
\end{equation}

We now provide explicit expressions for the components $D_{\parallel}$ and $D_{\perp}$ of this diffusion tensor in wave-vector space. Introducing a spherical coordinate system with a ${\bf z}$-axis that coincides instantaneously with the direction of the group velocity $\mathbf{v}_{g}$, we obtain

\begin{equation}\label{dij-spherical}
D_{ij}=\frac{\pi \, \overline{\omega}_{pe}^{4}}{4 \, \overline{\omega}^{2}} \int_0^\infty dq \, q^{2} \left . \frac{\widetilde{n}}{\overline{n}} \right \vert^{2}_{q} \, \int_{0}^{\pi} d\theta \, \sin \theta \int _{0}^{2 \pi} d\phi \, q_{i} \, q_{j} \, \delta (q \, v_{g} \, \cos \theta) \,\,\, .
\end{equation}
The non-diagonal components vanish by symmetry. The perpendicular diagonal components

\begin{equation}\label{dxx-dyy}
D_{xx} = D_{yy} = \frac{\pi \, \overline{\omega}_{pe}^{4}}{4 \, \overline{\omega}^{2}v_{g}} \int_0^\infty dq \, q^{3} \,\left . \frac{\widetilde{n}}{\overline{n}} \right \vert^{2}_{q} \, \int _{0}^{2 \pi} d\phi \, \cos^{2}\phi \int_{0}^{\pi} d\theta \, \sin^{3} \theta \, \delta (\cos \theta) \,\,\, ,
\end{equation}
which, with the change of variable $\mu=\cos \theta$, become

\begin{equation}\label{dxx-dyy-mu}
D_{xx}=D_{yy} = \frac{\pi^2 \, \overline{\omega}_{pe}^{4}}{4 \, \overline{\omega}^{2}v_{g}} \int_0^\infty dq \, q^{3} \left . \frac{\widetilde{n}}{\overline{n}} \right \vert^2_{q}
\int_{-1}^{1} d\mu \, (1-\mu^{2}) \, \delta (\mu) =\frac{\pi^{2} \, \overline{\omega}_{pe}^{4}}{4 \, \overline{\omega}^{2}v_{g}}  \int dq \, q^{3} \left . \frac{\widetilde{n}}{\overline{n}} \right \vert^2_{q}  \,\,\, .
\end{equation}
A similar calculation for the parallel component of the diffusion tensor gives

\begin{equation}\label{dzz}
D_{zz}=\frac{\pi^{2} \, \overline{\omega}_{pe}^{4}}{2 \, \overline{\omega}^{2}v_{g}} \int dq \, q^{3} \,\left . \frac{\widetilde{n}}{\overline{n}} \right \vert^{2}_{q} \, \int_{0}^{\pi} d\theta \, \cos^{2} \theta \, \sin \theta \, \delta (\cos \theta) =\frac{\pi^{2} \, \overline{\omega}_{pe}^{4}}{2 \, \overline{\omega}^{2}v_{g}} \int dq \, q^{3} \,\left . \frac{\widetilde{n}}{\overline{n}} \right \vert^{2}_{q} \, \int_{-1}^{1} d\mu \, \mu^{2} \, \delta (\mu) =0 \,\,\, .
\end{equation}
Overall, then, we have

\begin{equation}\label{dparallel-expression}
D_{\parallel} = 0,
\end{equation}
as anticipated, and

\begin{equation}\label{dperp-expression}
D_{\perp} = \frac{\pi^{2} \, \overline{\omega}^{4}_{pe}\int_0^\infty dq \, q^{3} \left . \frac{\widetilde{n}}{\overline{n}} \right \vert ^{2}_{q}}{4 \, \overline{\omega}^{2} \, v_{g} } \,\,\, .
\end{equation}

We next define the average magnitude of the relative density fluctuations

\begin{equation}\label{normalization}
\epsilon^{2}= 4 \pi \int d\Omega \,  dq \, q^2 \, \left . \frac{\widetilde{n}}{\overline{n}} \right \vert^{2}_{\mathbf{q},\Omega} \equiv
\left \langle \, \left ( \frac{\widetilde{n}}{\overline{n}} \right )^{2} \, \right \rangle \,\,\, ,
\end{equation}
and the average wavenumber (cm$^{-1}$) of the density fluctuations

\begin{equation}\label{average-q}
\langle \, q \, \rangle  \equiv \frac{4 \pi \int d\Omega \, dq \, q^{3} \left . \frac{\widetilde{n}}{\overline{n}} \right \vert ^{2}_{q,\Omega}}{\epsilon^{2}} \,\,\, .
\end{equation}
Using $\langle q \rangle$ and $\epsilon$, we obtain the more compact expression

\begin{equation}\label{dperp-qavg}
D_{\perp}=\frac{\pi}{16} \, \frac{\overline{\omega}^{4}_{pe} \, \langle q \rangle \, \epsilon^{2}}{\overline{\omega}^{2} \,  v_{g}}=\frac{\pi}{16} \, \frac{\langle q \rangle \, \epsilon^{2}}{c}\frac{\overline{\omega}^{4}_{pe}}{\overline{\omega}^{2} \,  \overline{\mu}_{r}} \,\,\, .
\end{equation}
We note that the effect of turbulence
on scattering enters as the product $\langle q \rangle \, \epsilon^{2}$ (cm$^{-1}$) in Equation~(\ref{dperp-qavg}), and that the quantity

\begin{equation}\label{nu-def-orig}
\nu=\frac{D_{\perp}}{k^{2}} = \frac{\pi}{16} \langle q \rangle \, \epsilon^{2} \, c \, \frac{\overline{\omega}_{pe}^{4}}{\omega^{4}
\overline{\mu}^{3}_{r}}
\end{equation}
has the meaning of a scattering frequency (s$^{-1}$), with a corresponding mean free path (cm)

\begin{equation}\label{mfp}
\lambda=\frac{v_{g}}{\nu}=\frac{\overline{\mu}_{r}c}{\nu} \,\,\, .
\end{equation}
These have the usual interpretations as the reciprocal of the characteristic time scale $\propto \langle q \rangle \,  c$, and length scale $\propto \langle q \rangle ^{-1}$, respectively, for 90$^o$ scattering due to the cumulative effect of many small angle deflections in wave-momentum space. Notice that the above derived coefficients $D_{\perp}$, $\nu$ and $\lambda^{-1}$, measuring the scattering efficiency, all behave as inverse positive powers of the refractive index $\overline{\mu}_{r} \leq 1$. When the emitted frequency $\omega$ is of the order of the plasma frequency, $\omega/\overline{\omega}_{pe} \sim 1$, the refractive index $\overline{\mu}_{r} = (1-\overline{\omega}_{pe}^{2}/\omega^{2})^{1/2}$ is close to zero. In such a case the scattering frequency $\nu$ diverges as $\overline{\mu}_{r}^{-3} = (1-\overline{\omega}_{pe}^{2}/\omega^{2})^{-3/2}$ and the mean free path approaches zero. On the other hand, when the emitted frequency $\omega$ is much larger than the plasma frequency, $\omega \gg \overline{\omega}_{pe}$, the refractive index $\overline{\mu}_{r}$ is close to unity and the scattering frequency decreases as $\omega^{-4}$. The weak $\overline{\mu}_{r}\sim 1$ scattering regime of radio sources was first considered by \cite{1952MNRAS.112..475C} and the results were generalized by \cite{1968AJ.....73..972H} to include the strong ($\overline{\mu}_{r}\sim 0$) scattering regime.

In summary, when the condition (\ref{static-condition}) is realized, the photon scattering operator~(\ref{dij-spherical}), expressed in a spherical coordinate system in ${\bf k}$-space, reduces to

\begin{equation}\label{scattering-operator}
\nu \left [ \frac{1}{\sin \theta}\frac{\partial }{\partial \theta}
\left ( \sin \theta \, \frac{\partial}{\partial \theta} \right ) + \frac{1}{\sin^{2} \theta} \,
\frac{\partial^{2}}{\partial \phi^{2}} \right ] \,\,\, ,
\end{equation}
which we recognize as the expression for the scattering operator in the Lorentz form \citep{1964pkt..book.....M, 1964PhFl....7..407K}.  Here, however, it describes not particle scattering, but rather photon diffusion on a sphere of constant radius in wave-vector space. The eigenfunctions of this diffusion operator are, of course, the spherical harmonics $Y_{\ell}^{m}(\theta,\phi)$. Moreover, since the average plasma frequency $\overline{\omega}_{pe}$ decreases with density and hence with heliospheric distance, the scattering frequency (\ref{nu-def-orig}) entering the scattering operator (\ref{scattering-operator}) is expected to be very large close to the intrinsic source of solar radio bursts, where $\omega \sim \overline{\omega}_{pe}$, and to become smaller as the photons propagate toward the observer, where $\omega \gg \overline{\omega}_{pe}$. This transition from strong to weak scattering can essentially be considered as a refraction effect, in which the group velocity $v_{g}$  of emitted photons increases as the photons propagate toward the observer.

So far, the description of photon transport has been very general; it accounts for the effect of refraction due to relatively smooth density variation and for the effect of scattering due to small-scale turbulent fluctuations.  It applies to both the strong and weak scattering regimes and allows for a transition between these two regimes as photons propagate away from the source.  We next consider some interesting special cases.

\section{Analysis of the small-angle scattering regime}\label{analysis}

We shall now proceed to gain further insight into the role of scattering in the Fokker-Planck equation~(\ref{fp-equation-f}) in the case where $D_{\parallel}=0$.  We set the large-scale refractive term to zero in Equation~(\ref{fp-equation-f}), effectively assuming that the plasma frequency, and therefore the refractive index, is constant within the transport region.  This implies that the absolute value $k$ of the wave-vector is a constant of motion and that $D_{\perp}(k) = {\rm const}$. We also consider the forward, small-angle limit, hence reducing the Fokker-Planck photon diffusion equation to a variant of Fermi's pencil beam equation \citep[unpublished courses at Chicago University; see][]{PhysRev.74.1534}.

Let us take a cartesian coordinate system with its origin at the center of the Sun and the $z$-axis pointing toward the observer (note that this $z$-axis is now fixed in space and not, as in the previous section, always parallel to the group velocity). We assume the source to be situated somewhere along the $z$-axis at a certain distance above the solar radius. The forward scattering regime corresponds to the limit of an angular spread sufficiently small that back scatter ($>90^\circ$) does not occur. We thus write $\mathbf{k} = (k_{z} + \delta~\!k_{z}) \, \mathbf{z} + \delta \mathbf{k}_{\perp}$, so that we are considering small deviations $\delta \mathbf{k}_{\perp}$ perpendicular to the emitted wave-vector $k_{z} \, \mathbf{z}$.  Since spherical diffusion in wave-vector space must preserve the value of $|\mathbf{k}|$, for each small perpendicular scattering increment $\delta k_{\perp}$, there must be an associated, higher order, decrement in the parallel wavevector $\delta k_{z} = O(\delta k_{\perp}^{2})$, given by

\begin{equation}\label{deltak_z}
\delta k_{z} \simeq \frac{1}{2} \, k_{z} \left ( \frac{\delta k_{\perp}}{k_{z}} \right )^{2} \,\,\, .
\end{equation}
This expression accounts for the statistical slowing-down of photons in the direction $\mathbf{z}$ of their initial emission, and hence for pulse broadening.  In this approximation, the Fokker-Planck equation reduces to

\begin{equation}\label{fp-eqn-small-scattering}
\frac{\partial N}{\partial t} +
\frac{c^{2}}{\omega} \, \mathbf{\delta k_{\perp}} \cdot \frac{\partial N}{\partial \mathbf{r_{\perp}}} + \frac{c^{2}}{\omega} \, k_{z} \left ( 1-\frac{1}{2} \left ( \frac{\delta k_{\perp}}{k_{z}} \right )^{2} \right ) \, \frac{\partial N}{\partial z} = D_{\perp} \, \frac{\partial ^{2}N}{\partial \, (\delta \mathbf{k}_{\perp})^{2}} \,\,\, ,
\end{equation}
describing respectively the effect of diffusion in $\delta \mathbf{k}_{\perp}$ (the right-hand side of this equation) on \emph{both} the spatial perpendicular spread $\mathbf{r}_{\perp}$ and on the parallel (to $z$) spread of scattered photons (the two last terms on the left-hand side of the equation). This equation is a generalization of Fermi's pencil beam equation in the sense that it includes the physics of slowing down in the parallel direction.  Although perpendicular dispersion and parallel slowing-down are driven by the same diffusion \citep[the perpendicular and parallel part of this Fokker-Planck equation are two standard equations in mathematical physics; see, e.g.,][]{Calin2009}, we shall now consider these two phenomena separately.

\subsection{Perpendicular dynamics and angular broadening}

Averaging the Fokker-Planck equation~(\ref{fp-eqn-small-scattering}) over the parallel spatial coordinate in space ($z$) and over one perpendicular coordinate (e.g., $k_{y}$) in wave-vector space gives

\begin{equation}\label{fp-eqn-1d}
\frac{\partial N}{\partial t} + \frac{c^{2}}{\omega} \, \delta k_{x} \, \frac{\partial N}{\partial x}
=D_{\perp} \, \frac{\partial ^2 N}{\partial (\delta k_x)^2} \,\,\, .
\end{equation}
Alternatively, this equation can be obtained from Equation~(\ref{fp-eqn-small-scattering}) by neglecting the $ O(\delta k_{\perp}^{2})$ transport term (hence discarding the effect of the statistical dispersion of the pulse in the parallel direction) and using the deterministic relation

\begin{equation}
z = \frac{c^{2}}{\omega} \, k_{z}\, t \,\,\, ,
\end{equation}
valid since $k_{z}$ is taken to be a constant of the motion. Equation (\ref{fp-eqn-1d}) is precisely Fermi's pencil beam equation describing the joint evolution of a photon cloud in both perpendicular space $x$ and wave-number space $\delta k_{x}$ as a function of time (or, equivalently, as a function of the parallel spatial coordinate $z$). This Fokker-Planck equation states that the perpendicular spatial coordinate $x$ of a scattered photon evolves as

\begin{equation}\label{dxbdt}
\frac{dx}{dt} =\frac{c^{2}}{\omega} \, \delta k_{x} \,\,\, ,
\end{equation}
while $\delta k_{x}$ diffuses at a rate given by

\begin{equation}\label{1}
\frac{d \langle \delta k_{x}^{2} \rangle}{dt}=2D_\perp \,\,\, ,
\end{equation}
a relation that can also be formally verified by taking the second moment of the Fokker-Planck equation. It follows from Equation~(\ref{dxbdt}) that

\begin{equation}\label{2}
\frac{d \langle x^{2} \rangle}{dt}=\frac{2c^{2}}{\omega} \langle \delta k_{x} \, x \rangle
\end{equation}
and

\begin{equation}\label{3}
\frac{d \langle \delta k_{x} \, x \rangle }{dt} = \frac{c^{2}}{\omega} \, \langle \delta k^{2}_{x} \rangle \,\,\, .
\end{equation}
Equations~(\ref{1}), (\ref{2}), and~(\ref{3}) form a closed set of moment equations, which combined give

\begin{equation}\label{4}
\frac{d^3 \langle x^2 \rangle}{dt^3} = \frac{2 c^2}{\omega} \, \frac{d^2 \langle \delta k_{x} \, x \rangle }{dt^2} = \frac{2 c^4}{\omega^2} \frac{d \langle \delta k_{x}^{2} \rangle}{dt}=\frac{4 c^4}{\omega^2} \, D_\perp \,\,\, ,
\end{equation}
with a similar equation for $\langle y^2 \rangle$.  Straightforward integration then gives

\begin{equation}\label{mean-square-r-perp}
\langle r_{\perp}^{2} \rangle = \langle x^{2} \rangle + \langle y^{2} \rangle = \frac{4}{3} \, \frac{c^{4}}{\omega^{2}} \, D_{\perp} \, t^{3} \,\,\, ,
\end{equation}
showing that the perpendicular spread of photons is faster ($\sim t^3$) than diffusive ($\sim t$) in time. Defining the angle

\begin{equation}\label{space-angle-psi}
\psi \simeq \frac{\sqrt{x^2 + y^2}}{z} \,\,\, ,
\end{equation}
we obtain, in the small angle approximation, the diffusive behavior in space:

\begin{equation}\label{small-angle-diffusion}
\langle \psi^{2} \rangle = D_{\psi} \, z \,\,\, ,
\end{equation}
where

\begin{equation}\label{dpsi-expression}
D_{\psi} = \frac{\pi}{12} \,  \langle q \rangle \, \epsilon^{2} \, \frac{\overline{\omega}_{pe}^{4}} {\overline{\omega}^{4} \, \overline{\mu}_r^{4}}
\end{equation}
is the angular diffusion coefficient (cm$^{-1}$) describing the linear increase in $\langle \psi^{2} \rangle$ as a function of the distance $z = v_{g} t $ from the source. Notice that $D_{\psi}$ inherits its physical dimension (cm$^{-1}$) solely from $\langle q \rangle$, the reciprocal of the characteristic length-scale of the density fluctuations. Following \cite{1952MNRAS.112..475C},  we also define the angles

\begin{equation}\label{defangle}
\Psi_{1} \equiv\frac{dx}{dz} = \frac{\delta k_{x}}{k_{z}} \,\,\, , \qquad \Psi_{2} \equiv\frac{dy}{dz}=\frac{\delta k_{y}}{k_{z}} \,\,\, .
\end{equation}
We stress that these quantities represent angles in the wave-vector domain as opposed to the angle $\psi$ previously defined in the spatial domain.  The total mean square deflection angle is thus given by

\begin{equation}\label{psi2}
\langle \Psi^{2} \rangle =\frac{\langle \delta k^{2}_{x}\rangle }{k^{2}_{z}}+\frac{\langle \delta k^{2}_{y}\rangle }{k^{2}_{z}}=\frac{2 \langle \delta k^{2}_{x} \rangle}{k^{2}_{z}} \,\,\, .
\end{equation}
Taking the time derivative of Equation (\ref{psi2}), and since $k_{z}$ is a constant of motion, we obtain

\begin{equation}\label{dpsi3}
\frac{d \langle \Psi^{2} \rangle }{dt} =\frac{2}{k^{2}_{z}}\frac{d \langle \delta k^{2}_{x} \rangle}{dt}=\frac{4D_{\perp}}{k^{2}_{z}}
\end{equation}
and hence

\begin{equation}
\frac{d\langle \Psi^{2}\rangle }{dt} = 4\, \nu \,\,\, ,
\end{equation}
where $\nu$ is the scattering frequency defined in Equation~(\ref{nu-def-orig}). Finally using $dz=v_{g} \, dt$ we have

\begin{equation}\label{dpsi2}
\frac{d\langle \Psi^{2}\rangle }{dz} = \frac{4}{\lambda} = 3 \, D_{\psi} \,\,\, .
\end{equation}
This is essentially the result of \cite{1952MNRAS.112..475C} \citep[see also][]{1968AJ.....73..972H} which is here readily recovered from Fermi's pencil-beam equation \citep{PhysRev.74.1534}. The angular diffusion coefficients are here expressed in term of the scattering mean free path. We shall now derive the joint PDF in space and wave-number space (or angular space), from which the result~(\ref{dpsi2}) originates.

Introducing the dimensionless wave-number, time, and space variables

\begin{equation}\label{nd-variables-1}
\zeta=\frac{\delta k_{x}}{\sqrt{2} \, k_{z}}\, ; \qquad
\tau = \nu \, t \, ; \qquad X = \frac{\omega \, \nu}{\sqrt{2} \, c^2 \, k_{z}} \, x \,\,\, ,
\end{equation}
Equation~(\ref{fp-eqn-1d}) reduces to the following differential equation for the probability distribution function $P(X, \zeta, \tau)$:

\begin{equation}\label{fp-eqn-1d-red}
\frac{\partial P}{\partial \tau} = L_K \, P  \,\,\, ,
\end{equation}
where the \cite{10.2307/1968123} operator

\begin{equation}\label{ko}
L_K = \frac{1}{2} \, \frac{\partial ^{2}}{\partial \zeta^{2}} - \zeta \, \frac{\partial}{\partial X} \,\,\, .
\end{equation}
The solution of Equation~(\ref{fp-eqn-1d-red}) corresponding to the initial condition $P(X,\zeta,\tau=0)=\delta(X) \, \delta (\zeta)$, explicitly derived by, e.g., \citet{Calin2009}, is

\begin{equation}\label{p-solution}
P(X,\zeta,\tau)=\frac{\sqrt{3}}{\pi \tau^{2}} \, \exp{ \left ( -\frac{\zeta ^{2}}{2\tau} - \frac{6}{\tau^{3}} \, \left ( X-
\frac{\tau}{2} \, \zeta \right )^{2} \right )} \,\,\, ,
\end{equation}
showing that the joint PDF characterizing the perpendicular dynamics of photons is Gaussian despite the spatial dispersion of photons being superdiffusive\footnote{the \cite{10.2307/1968123} operator (\ref{ko}) entering Fermi's pencil beam equation appears to have been introduced originally in fluid mechanics; it also forms the basis of the \cite{1959AdGeo...6..113O} model for the phenomenon of Richardson dispersion, i.e., the explosive separation of fluid particles and field lines in turbulent flows \citep{2013Natur.497..466E, PhysRevLett.117.095101}, and more simply for the phenomenon of shear dispersion of a passive tracer \citep{2005PhPl...12a4503G}.}, $\langle x^{2} \rangle(t)\sim t^{3}$.
As we shall see in the next section, however, the PDF characterizing the parallel dynamics is far from Gaussian.

\subsection{Longitudinal dynamics and pulse profile}\label{pulse-profile-analysis}

We now turn to the analysis of the spatial dispersion along the $z$-axis which results from parallel slowing-down, i.e., from the $O(\Psi^{2})$ decrease in parallel wave number $\delta k_{z}\simeq \frac{1}{2} \, k_{z} \Psi^{2}$, that results from angular broadening in the small angle $\Psi\ll 1$ limit. Such a statistical slowing-down will manifest itself in variable time delays, depending on the different random paths taken by the photons, and hence in a time-broadening of the observed profile. To study this aspect of photon propagation, we now average the Fokker-Planck equation~(\ref{fp-eqn-small-scattering}) over the perpendicular spatial degrees of freedom:

\begin{equation}\label{fp-eqn-time-profile}
\frac{\partial N}{\partial t} +
\frac{c^{2}}{\omega} \, k_{z} \, \left ( 1 - \frac{1}{2} \left ( \frac{ \delta k_{x}}{k_{z}} \right )^{2}
-\frac{1}{2} \left (\frac{ \delta k_{y}}{k_{z}} \right )^{2} \right ) \frac{\partial N}{\partial z}
= D_{\perp} \left ( \frac{\partial ^{2}N}{\partial (\delta k_{x})^{2}}+\frac{\partial ^{2}N}{\partial (\delta k_{y})^{2}} \right ) \,\,\, .
\end{equation}
This equation states that due to scattering, a photon lags behind an unscattered photon propagating at the speed $v_{g}$ by an amount $\delta z(t)$ satisfying the equation

\begin{equation}\label{d-deltaz-dt}
\frac{d \, \delta z}{dt}= \frac{v_{g}}{2} \left [ \left ( \frac{ \delta k_{x}}{k_{z}} \right )^{2}+
\left ( \frac{ \delta k_{y}}{k_{z}} \right )^{2} \right ]= \frac{1}{2} v_{g} \, \Psi^{2} \,\,\, .
\end{equation}
This means that a photon is delayed at the observing point relative to an unscattered photon by an amount $\delta t (t)$ satisfying

\begin{equation}\label{delta-t-def}
\frac{d \delta t}{dt}= \frac{1}{2} \left [ \left ( \frac{ \delta k_{x}}{k_{z}} \right )^{2}+
\left ( \frac{ \delta k_{y}}{k_{z}} \right )^{2} \right ]= \frac{1}{2} \, \Psi^{2} \,\,\, .
\end{equation}
Taking the time derivative of Equation~(\ref{delta-t-def}), again using that $k_{z}$ is a constant of motion, and averaging the result, we obtain

\begin{equation}\label{dtdelta}
\frac{d^{2} \langle \delta t\rangle}{dt^{2}}=\frac{1}{2}\frac{d \langle \Psi^{2}\rangle }{dt} = 2 \, \nu \,\,\, ,
\end{equation}
or equivalently

\begin{equation}\label{dzdelta}
\frac{d^{2} \langle \delta t\rangle}{dz^{2}}=\frac{D_{\Psi}}{2 \, \overline{\mu}_{r} \, c} = \frac{2}{\overline{\mu}_{r}c} \, \frac{1}{\lambda} \,\,\, .
\end{equation}
These equations describe the rates of increase of the pulse width $\langle \delta t \rangle$ as a function of either time $t$ or distance $z$ from the source. Straightforward integration yields

\begin{equation}\label{nu-def}
\langle \delta t \rangle = \, \nu \, t^{2}=\frac{\nu}{c^{2} \, \overline{\mu}^{2}_{r}} \, z^{2} \,\,\, .
\end{equation}
We note that this behavior is quadratic (i.e., again faster than diffusive) in both space and time. The delay time $\delta t$ is inherently a random quantity and we shall now discuss the form of the associated probability distribution function (PDF) $P(\delta t, t)$, i.e., the PDF of the delay time $\delta t$ at time $t$. We define the dimensionless wave-number, time, and delay time by

\begin{equation}\label{nd-variables-2}
\zeta=\frac{\delta k_{x}}{\sqrt{2} \, k_{z}} \, ; \qquad
\tau = \nu \, t \, ; \qquad T = \nu \, \delta t \,\,\, .
\end{equation}
The problem then reduces at computing the PDF $Q(T,\tau)$ of the delay time $T$ given by

\begin{equation}\label{dT-dtau}
\frac{dT}{d\tau}= \zeta^{2} \,\,\, ,
\end{equation}
where $\zeta(\tau)$ represents a diffusion process. Such a problem is also the domain of Brownian functionals \citep{2005cond.mat.10064M}; here it consists in computing $T(\tau)=\int_{0}^{\tau} ds \, \zeta^{2}(s)$, which is the sum of the square of the Brownian motion. The evolution of the joint PDF $Q(T, \zeta, \tau)$ is governed by the Fokker-Planck equation

\begin{equation}\label{fp-eqn-time}
\frac{\partial Q}{\partial \tau} + \zeta^{2} \, \frac{\partial Q}{\partial T}=\frac{1}{2} \, \frac{\partial ^{2}Q}{\partial \zeta^{2}} \,\,\, .
\end{equation}
Since $T$ is positive, we can use the Laplace transform

\begin{equation}\label{l-transform-Q}
{\widehat Q}(u,\zeta,\tau) = \int _{0}^{\infty}e^{-uT} \, Q(T, \zeta, \tau) \, dT \,\,\, ,
\end{equation}
and the transform of Equation~(\ref{fp-eqn-time}) is

\begin{equation}\label{fp-eqn-time-red}
\frac{\partial\widehat  Q}{\partial \tau} = L_H \, {\widehat Q} \,\,\, .
\end{equation}
Here the Hermite operator

\begin{equation}\label{hermite-operator}
L_H = \frac{1}{2} \, \frac{\partial ^{2}}{\partial \zeta^{2}} - u \, \zeta^{2} \,\,\, ,
\end{equation}
which has as eigenfunctions the Hermite polynomials $H_{n}((2u)^{1/4} \, \zeta)$. The Green function solution of this equation, corresponding to the initial condition $\widehat Q(\tau=0, \zeta)=\delta(\zeta-\zeta_{0})$, is the Mehler kernel

\begin{equation}\label{mel}
\hat Q(u,\zeta,\zeta_{0},\tau)=\frac{1}{\sqrt{2\pi \tau}}\sqrt{\frac{\sqrt{2u}\tau}{\sinh \sqrt{2u}\tau}} \, \exp{ \left \{ {-\frac{1}{2\tau} \, \frac {\sqrt{2u}\tau}{\sinh \sqrt{2u}\tau} \, [ \, (\zeta^{2}+\zeta_{0}^{2}) \,
\cosh(\sqrt{2u}\tau) - 2 \, \zeta \, \zeta _{0} \, ]} \right \} } \,\,\, ;
\end{equation}
an explicit derivation of this result, which stems for the moment generating function of the joint PDF $Q(T,\zeta, \zeta_{0}, \tau)$, can be found in, e.g., \cite{Calin2009}.  In the following, we provide an analytical approximation for the marginal PDF $Q(T,\tau)$, giving the delay time $T$ of a photon at time $\tau$ independently of the value of $\zeta$ involved in this delay, i.e., we concentrate on the arrival time profile only. The (normalized) wave-number variable $\zeta$ is zero initially (recall that the photon is emitted along $z$ initially) so we set $\zeta_{0}=0$ in the expression (\ref{mel}) and perform the average over $\zeta$:

\begin{equation}\label{qhat-transform}
\widehat Q(u,\tau)=\int d\zeta \, {\widehat Q} (u,\zeta,\zeta_{0}=0,\tau) \,\,\, ,
\end{equation}
which yields

\begin{equation}\label{qhat-expression}
\widehat{Q}(u,\tau) = \left ( \frac{1}{\cosh (\tau\sqrt{2u}) } \right )^{1/2}
\end{equation}
as the Laplace transform of the PDF $Q(T,\tau)$ of the delay time $T$. Including the contribution to the delay time due to diffusion in the additional degree of freedom $\delta k_{y}$ (which is independent of the diffusion in $\delta k_{x}$ but has the same Gaussian statistics), and since the Laplace transform of the sum of two independent random variables is the product of their Laplace transforms, we find that the Laplace transform of $P(T,\tau)$ is

\begin{equation}\label{phat-transform}
{\widehat P}(u,\tau) = {\widehat Q}^2(u,\tau) = \frac{1}{\cosh (\tau\sqrt {2u}) }  \,\,\, .
\end{equation}
The Laplace inversion with respect to $u$ of this expression is made by first expanding it:

\begin{equation}\label{ltp}
{\widehat P}(u,\tau) = \frac{1}{\cosh (\tau\sqrt {2u}) } = \frac{2 \, e^{-\tau \sqrt{2u} }}{1+e^{-2\tau\sqrt{2u} }}= 2\sum_{n=0}^{\infty}(-1)^{n}e^{-(2n+1)\tau\sqrt{2u} }
\end{equation}
and then inverting this term by term, using the Levy formula

\begin{equation}\label{levy}
\int_{0}^{\infty} \frac{a}{\sqrt{2\pi T^{3}}} \, e^{-a^{2}/2T} \, e^{-uT} \, dT = e^{-a\sqrt{2u}} \,\,\, .
\end{equation}
This gives

\begin{equation}\label{P-t-tau-expression}
P(T,\tau)=2 \, \sum_{n=0}^{\infty} \, (-1)^{n} \,
\frac{ \left ( 2n + 1 \right ) \tau}{\sqrt{2\pi T^{3}}} \, \exp \left [ -\frac{ \left ( 2n + 1 \right)^{2}  \tau^{2}}{2 \, T} \right ] \,\,\, .
\end{equation}
Returning to the original variables, we obtain the expression for the PDF of the delay time $\delta t$ as a function of time $t$:

\begin{equation}\label{numerical-expression}
P(\delta t,t) = \frac{2\nu t}{\sqrt{2\pi \, \nu^{3} \, \delta t^{3}}} \, \sum_{n=0}^{\infty} \, (-1)^{n} \,
 \left ( 2n + 1 \right ) \exp \left [ -\frac{ \left ( 2n + 1 \right)^{2} \nu  \, t^{2}}{ \delta t} \right ] \,\,\, .
\end{equation}

\begin{figure}[pht]
\centering
\includegraphics[width=0.7\textwidth]{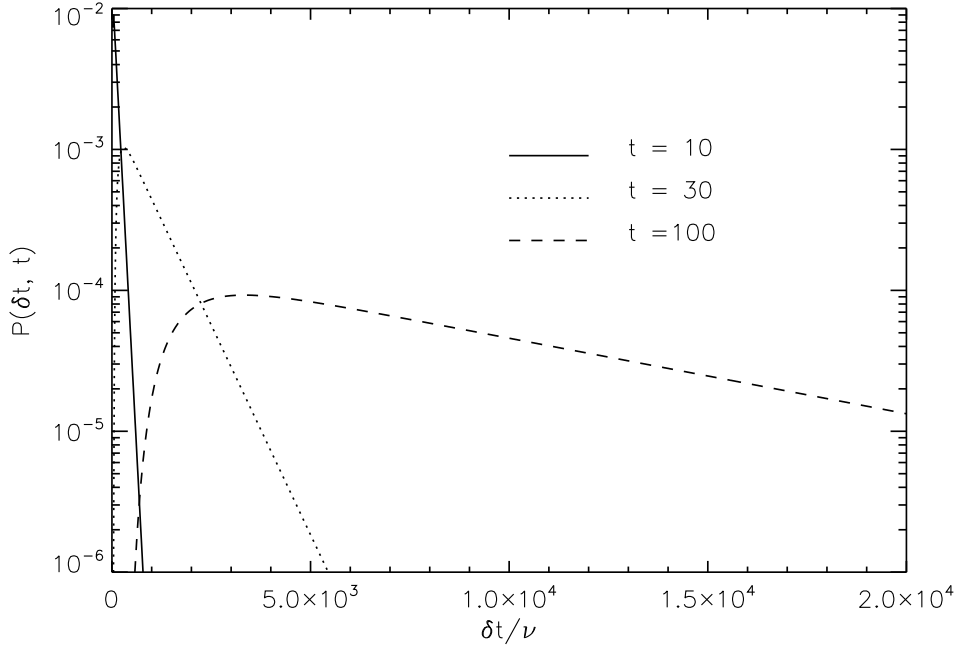}
\caption{Evaluation of the expression~(\ref{numerical-expression}) for $P(\delta t,t)$.  The form of $P(\delta t,t)$ is characterized by a rapid rise followed by an exponential decay, with timescale $\nu \, t^2$ (see text).}\label{numerical-solution}
\end{figure}

Figure~\ref{numerical-solution} shows the numerical evaluation of this expression, where the sum over $n$ has been taken up to $n=30$ (adding additional terms has a negligible effect on the solution).  We see that the form of $P(\delta t,t)$ is characterized by a steep rise followed by an exponential decay. For small values of $\delta t/\nu \ll t^2$, the successive terms in this series converge rapidly due to the $(2n+1)^2$ term in the exponential, and indeed numerical evaluation shows that the first term in the series:

\begin{equation}\label{first-term-in-series}
P(\delta t,t)=\frac{2 \, \nu \, t}{\sqrt{2\pi \nu^{3}\delta t^{3}}} \, \exp \left [ - \frac{ \nu  \, t^{2}}{ \delta t} \right ]
\end{equation}
is a good representation of the whole sum.  This accounts for the sharp initial increase in $P(\delta t,t)$ and for the start of the decay. At the other extreme, for large $T$, we can approximate the Laplace transform ${\hat P}(u,\tau)$ (Equation~(\ref{ltp})) by its small $u$ approximation:

\begin{equation}\label{small-u-approx}
{\widehat P}(u,\tau) = \frac{1}{\cosh (\tau \sqrt{2u})} \simeq \frac{1}{1 + \tau^2 u} \,\,\, .
\end{equation}
This has inverse Laplace transform

\begin{equation}\label{large-T}
P(T,\tau) = \frac{1}{\tau^2} \, e^{-T/\tau^2} \,\,\, ,
\end{equation}
showing that the PDF of the delay time $T$ has an exponential behavior at large $T$.  In terms of the original variables, the exponential decay time scale is given by

\begin{equation}\label{tau-d}
\tau_d = \nu \, t^{2},
\end{equation}
the decay time of the radio pulse is thus equal to the average delay time $\langle \delta t \rangle$ (Equation~(\ref{nu-def})), and hence

\begin{equation}\label{tau-d-omega}
\tau_d = \langle \delta t \rangle=\frac{\pi}{16} \, \frac{\langle q \rangle \, \epsilon^{2} \, L^2}{c} \, \frac{\overline{\omega}^{4}_{pe}}{\omega^{4} \, \overline{\mu}_{r}^{5}} \,\,\, ,
\end{equation}
where $L$ is the distance from the source. The model therefore predicts that the decay time $\tau_{d} \propto 1/\overline{\omega}^{4}$ in the weak scattering regime when $\omega \gg \overline{\omega}_{pe}$  and diverges like $\overline{\mu}_{r}^{-5}=(1-\overline{\omega}^{2}_{pe}/\omega^{2})^{-5/2}$ in the strong scattering regime when $\omega \sim \overline{\omega}_{pe}$. We shall discuss the consistency of this result (and the approximations involved in deriving it) with observations in Section~\ref{discussion}.

\section{Analysis of the diffusive regime}\label{diffusive-analysis}

Angular scattering in wave-vector space tends, over time, to isotropize the photon distribution. When $t\gg \nu^{-1}$, any given photon in the ensemble now undergoes a spatial diffusion (a Brownian motion in space), which must itself be isotropic if the scattering process is isotropic. For such an isotropic diffusion process we can, without loss of generality, consider the dynamics along the $z$-direction only, described by

\begin{equation}\label{pplp}
\frac{\partial N}{\partial t}
+\frac{c^{2}}{\omega} \, k\cos\theta \cdot \frac{\partial N}{\partial z}=
\nu \left [ \frac{1}{\sin \theta} \, \frac{\partial }{\partial \theta}
\left ( \sin \theta \, \frac{\partial N}{\partial \theta} \right ) \right ] \,\,\, .
\end{equation}
With $\mu=\cos \theta$, this reads

\begin{equation}\label{fp-diffusive-mu}
\frac{\partial N}{\partial t}
+\mu \, v_{g} \, \frac{\partial N}{\partial z}=
\nu \left [ \frac{\partial }{\partial \mu}( 1-\mu^{2})\frac{\partial N}{\partial \mu} \right ] \,\,\, .
\end{equation}
Expanding the distribution function in Legendre polynomials:

\begin{equation}\label{legendre-expansion}
N=N_{0}(z,t)+\mu \, N_{1}(z,t) + \cdots \,\,\,
\end{equation}
we obtain \citep[cf.][]{2017ApJ...835..262B} the relation

\begin{equation}\label{flux-gradient-relation}
q_{z}=\kappa \, \frac{\partial N_{0}}{\partial z}
\end{equation}
between the spatial flux of photons along $z$ and the spatial gradient of the distribution.  The constant of proportionality $\kappa$, the spatial diffusivity of radio photons, is

\begin{equation}\label{kappa-def}
\kappa=\frac{1}{3} \, \frac{v_{g}^{2}}{\nu} = \frac{1}{3} \, \frac{\overline{\mu}_{r}^{2} \,c^{2}}{\nu}=\frac{1}{3} \, \lambda \, \overline{\mu}_{r} \, c
\end{equation}
(note that a high value of the scattering frequency leads to efficient scattering and hence a low value of the spatial diffusivity  $\kappa$). Since both the mean free path $\lambda$ and the refractive index $\overline{\mu}_{r}$ approach zero close to the source, the spatial diffusivity of photons is very small there. For isotropic scattering, such a relation must hold also in the directions $x$ and $y$, and therefore in three dimensions the isotropic part of the distribution of photons evolves according to the diffusion equation

\begin{equation}\label{diffusion-equation-3d}
\frac{\partial N_{0}}{\partial t}=\frac{\partial }{\partial \mathbf{r}} \cdot \left ( \kappa \, \frac{\partial N_{0}}{\partial \mathbf{r}} \right )  \,\,\, .
\end{equation}

\subsection{Pulse profile}

To determine the temporal profile of the radio burst in the diffusive regime of photon transport we now draw an analogy with the first-passage distributions of a Brownian motion \citep{2007gfpp.book.....R} outside a given spatial domain. We consider turbulent scattering to be important only inside a sphere of radius $R_{st} < 1$~AU centered at the source position. Note that since scattering is operating only inside this domain, the photon can only exit once, i.e., it cannot re-enter the spherical domain. Thus there is no distinction between the first passage time and the exit time. Due to spatial diffusion, the exit time $\delta t$ of a photon at $r=R_{st}$ is a random quantity and we identify (up to a geometrical dilution factor) the exit-time distribution $P(\delta t)$ with the intensity-time profile of the burst. Since diffusion is assumed to be isotropic, the PDF of the exit time must be independent of the particular exit point on this sphere. Therefore, the distribution of photon exit times in the neighborhood of a particular point on the surface of the sphere is, by symmetry, the same as $P(\delta t)$. It follows that the pulse width recorded  by a distant observer, say at $r=R_{1AU}$, is of the order of the diffusion time of photons over a radius $R_{st}$. This statement corresponds to Equation~(67) in \citet{1999A&A...351.1165A}, where a constant of proportionality is left undetermined. In the spherical case \citep{1998PhRvE..57.3937B}, it can be shown that

\begin{equation}\label{expectation-deltat}
\langle \delta t \rangle = \frac{R_{st}^{2}}{2\kappa} \,\,\, .
\end{equation}
The PDF $P(\delta t)$ can also be computed following the probabilistic method of Kac \citep[see][]{1998PhRvE..57.3937B}, wherein it is shown that the Laplace transform of the full PDF $P(\delta t)$ of the exit time of diffusing photons is given by

\begin{equation}\label{pdf-of-exit-time}
{\cal L}[P(\delta t)] \equiv \widehat{P}(u) = \frac{1}{\cosh \sqrt{u \langle \delta t \rangle }}
\end{equation}
\citep[see Equations (3.9) and (3.32) in][]{1998PhRvE..57.3937B}. Hence this model for the intensity profile in the diffusive regime gives a similar result to the small-angle regime: the pulse widths and decay times in both the small-angle and diffusive regimes have the same scaling with frequency $\omega$.  However, in the diffusive regime, the exponential decay time is the diffusion time over the scattering length $R_{st}$.

\subsection{Angular broadening}

If diffusion causes photons to be emitted isotropically from any given point on a sphere of radius $R_{st}$, this radius also determines the apparent size of the scattered image of the true source, situated inside the scattering region, as seen by an observer. The apparent angular size of the source observed at, say, 1~AU has then a purely geometric meaning, independent of the transport coefficient and hence on the level of density fluctuations within the scattering region; it is given by

\begin{equation}\label{apparent-angular-size-at-1au}
\psi \sim \frac{R_{st}}{1 \, {\rm AU}} \,\,\, ,
\end{equation}
when  $\psi \ll 1$.  This result corresponds to the statement in Figure~8 of \citet{1999A&A...351.1165A}, showing the saturation of the angular broadening at a value independent of the scattering frequency as the intensity of scattering is increased, i.e., when $\nu\rightarrow \infty$ and the diffusive regime of transport is reached inside the scattering region\footnote{This last point is in fact rather intuitive, as can be seen by considering another example: radiation from the Sun. The intrinsic source of photons is the core, and if photon production there were to stop abruptly, the Sun would continue to emit light for the thousands of years it takes for photons to scatter inside the Sun before exiting the last effective scattering surface, i.e., the photosphere. However, increasing the scattering frequency inside the Sun would not change the size of the photosphere, the scattered image of the photon production region. Thus, while increasing the scattering frequency will increase the arrival delay time scale, it will not increase the apparent source size.}.

\section{Fermi acceleration of radio photons near the intrinsic source}\label{fermi-acceleration}

When the scattering centers were considered as fixed in space, parallel diffusion in momentum space could be neglected (cf. Equations~(\ref{cfn}) and~(\ref{spn})).  However, scattering of photons at multiple randomly located sites along their paths may result in a Fermi-type parallel diffusion in wave-vector space.  Analysis of this phenomenon requires a statistical treatment of diffusion in wave-vector space which may in turn, as we have seen in Section~\ref{diffusive-analysis} above, lead to diffusion in space when isotropization is sufficiently strong. The domain of validity of the elastic approximation to wave momentum diffusion is

\begin{equation}\label{domain-elastic-approximation}
v_{\Phi} \equiv \frac{\Omega}{q} \ll v_g \,\,\, ,
\end{equation}
where $v_{\Phi}$ is the characteristic ``speed'' of the density fluctuations. In a tenuous plasma where $v_{g}$$\sim$$c$, or in the weak scattering regime $\overline{\mu}_{r}\sim 1$, this is always satisfied; however, this assumption will break down close to the critical level ($\omega=\omega_{pe}$) where the group velocity vanishes and where scattering is strong; here we must explicitly take into account the time dependence of the fluctuations. For such conditions the parallel part of the quasilinear diffusion coefficient (cf. Equation~(\ref{dzz}); repeated here) is

\begin{eqnarray}\label{parallel-part-of-dij}
D_{\parallel} & = & \frac{\pi^{2} \, \overline{\omega}_{pe}^{4}}{2 \, \overline{\omega}^{2}v_{g}} \int dq \, q^{3} \,\left . \frac{\widetilde{n}}{\overline{n}} \right \vert^{2}_{q} \, \int_{0}^{\pi} d\theta \, \cos^{2} \theta \, \sin \theta \, \delta  \left ( \cos \theta-\frac{\Omega(\mathbf{q})}{q \, v_{g}} \right ) \cr
& = & \frac{\pi^{2} \, \overline{\omega}_{pe}^{4}}{2 \, \overline{\omega}^{2}v_{g}} \int dq \, q^{3} \,\left . \frac{\widetilde{n}}{\overline{n}} \right \vert^{2}_{q} \, \int _{-1}^{+1}
d\mu \, \mu^{2} \, \delta \left ( \mu -\frac{\Omega(\mathbf{q})}{q \, v_{g}} \right ) \,\,\, .
\end{eqnarray}
This no longer vanishes and is responsible for statistical Fermi-type acceleration of photons. We shall consider two specific examples involving two types of waves likely involved in the mechanism of plasma emission: ion-sound waves and Langmuir waves.

\subsection {Ion-sound waves}

The generation of ion-sound waves is a natural product of the streaming of a distribution of fast electrons past a background Maxwellian distribution of stationary ions \citep[see, e.g.,][]{1980panp.book.....M}.  The role of ion-sound waves has been invoked to explain numerous characteristics of accelerated electrons in solar flares \citep[see, e.g.,][]{1979ApJ...228..592B}, and it is thus reasonable to expect a significant level of ion-sound turbulence also in the radio wave propagation region. Further, scattered radio waves are often observed in the ionosphere to be Doppler-shifted by an amount consistent with a velocity of the order of the ion-sound speed (i.e., $\sim$$\pm \, C_s$) \citep[e.g.,][]{2008AnGeo..26.2435S,2013JASTP.105..299M}. It is therefore natural to consider the influence of ion-sound waves on the propagation of radio waves in the active solar corona. The dispersion relation for ion-sound waves is

\begin{equation}\label{dispersion-relation-ion-sound}
\Omega(\mathbf{q}) = \pm \, q \, C_{s} \,\,\, .
\end{equation}
Inserting this in Equation~(\ref{parallel-part-of-dij}), and using Equations~(\ref{normalization}) and~(\ref{average-q}), gives the associated parallel component of the diffusion tensor:

\begin{equation}\label{parallel-diffusion-ion-sound}
D_{\parallel} = \frac{\pi}{8} \, \frac{\overline{\omega}_{pe}^{4}}{\overline{\omega}^{2}} \, \frac{\langle q \rangle \, \epsilon^{2}}{v_{g}} \,
\left ( \frac{C_{s}}{v_{g}} \right )^{2} \,\,\, .
\end{equation}
Comparing with the perpendicular diffusion coefficient (Equation~(\ref{dperp-qavg})), we find that

\begin{equation}\label{ratio-dparallel-dperp}
\frac{D_{\parallel}}{D_{\perp}} = 2\left ( \frac{C_{s}}{v_{g}} \right )^{2}\,\,\, ,
\end{equation}
showing that when $v_{g} \gg C_{s}$ parallel diffusion is only a small contribution to the full diffusion tensor, which is still dominated by angular scattering.  However, when $v_{g} \simeq C_{s}$ longitudinal Fermi acceleration (i.e., change in energy) is of the same order of (in fact twice) the angular scattering rate. Such considerations can be extended to turbulence produced by other low-frequency plasma modes such as Alfv\'en waves which become compressive at small-scales \citep{2009ApJS..182..310S, 2009PhPl...16f4503B, 2010A&A...519A.114B, 2010PhPl...17f2308B, 2017SoPh..292..117L}.

\subsection{Langmuir waves}

In the same manner, we can investigate the effect on the parallel diffusion of radio waves of a high-level of Langmuir waves in the propagation region.  Langmuir waves have the dispersion relation

\begin{equation}\label{dispersion-relation-langmuir-waves}
\Omega(\mathbf{q})=\omega_{pe} \,\,\, ,
\end{equation}
and substituting this in Equation~(\ref{parallel-part-of-dij}) gives the corresponding parallel component of the diffusion tensor:

\begin{equation}\label{dparallel-langmuir-waves}
D_{\parallel}=\frac{\pi^{2} \, \overline{\omega}_{pe}^{4}}{2 \, \overline{\omega}^{2}v_{g}} \int dq \, q^{3} \,\left . \frac{\widetilde{n}}{\overline{n}} \right \vert^{2}_{q} \, \int _{-1}^{+1} d\mu \, \mu^{2}\delta \left ( \mu - \frac{\omega_{pe}}{q \, v_{g}} \right )= \frac{\pi^{2}}{2} \, \frac{\overline{\omega}_{pe}^{6}}{\overline{\omega}^{2}} \, \frac{1}{v^{3}_{g}}\int dq \, q
\left . \frac{\widetilde{n}}{\overline{n}} \right \vert ^{2}_{q} \,\,\, .
\end{equation}
This shows that the statistical acceleration effect becomes large close to the source of emission, particularly for fundamental emission which has its group velocity $v_{g}$ very close to zero.  Physically, photons are so slow near the source of radio-bursts that they can ``feel'' the temporal variation of the fluctuations in addition to the spatial variation.

\subsection{Double-diffusion equation for photons}

Overall, close to the source the combined effects of strong angular scattering and Fermi acceleration are described by the double-diffusion equation

\begin{equation}\label{double-diffusion}
\frac{\partial N_{0}}{\partial t} = \frac{\partial }{\partial \mathbf{r}} \cdot \left ( \kappa \, \frac{\partial N_{0}}{\partial \mathbf{r}} \right ) + \frac{1}{k^{2}} \, \frac{\partial
}{\partial k} \left ( k^{2}D_{\parallel}\frac{\partial N_{0}}{\partial k} \right ) \,\,\, ,
\end{equation}
describing simultaneous diffusion in both space and in energy space. Notice that the spatial diffusion coefficient $\kappa$ explicitly depends on the value of the perpendicular diffusion coefficient $D_{\perp}$ in wave-vector space, i.e., $\kappa = \kappa(D_{\perp})$. In the classification scheme provided by \cite{2012ApJ...754..103B}, such a mechanism for stochastic acceleration, here of radio photons, is said to be resonant.

\section{Summary and Conclusions}\label{discussion}

So far, most analytical (and numerical) studies of scattering of radio waves in the corona and in interplanetary space have been based on the works by \cite{1952MNRAS.112..475C} and \cite{1968AJ.....73..972H}, which are both based on a perturbation of the eikonal equation. Other methods include the parabolic wave equation \citep{1975ApJ...196..695L,1975ApJ...201..532L,1995ApJ...439..494B,1998ApJ...506..456C}, which accounts for the effect of wave diffraction (the ``splitting'' of photon trajectories) in addition to that of refraction and scattering, but the parabolic wave equation is limited to the forward propagation regime and thus cannot account for the diffusive regime of photon transport.

In this work we have presented a comprehensive analytical study of scattering of radio waves in fluctuating plasmas based on the Fokker-Planck equation; this approach serves as a unified framework for modeling the propagation of radio bursts \citep[and also for modeling the propagation of radio-waves in interstellar space that originate from, e.g., pulsars; see][]{1978ppim.book.....S}. We have used the quasilinear theory of wave scattering \citep{1967PlPh....9..719V} to obtain a general expression for the diffusion tensor of photons in wave-vector space, where both spatial and temporal variations of the turbulent plasma fluctuations are considered. The latter generally involves two coefficients, $D_{\perp}$ and $D_{\parallel}$, respectively describing diffusion in angle and in the modulus of the wave vector {\bf k}, i.e., in the directions respectively perpendicular and parallel to the wave-vector. In the limit where density fluctuations can be considered quasi-static, corresponding to fast propagation of the electromagnetic radiation, $D_{\perp} \gg D_{\parallel}$, the scattering operator in the Fokker-Planck equation takes the Lorentz form, and diffusion in wave-vector space occurs at a constant modulus $\vert {\bf k} \vert$.

Neglecting the effect of refraction (i.e., effectively assuming a constant $\vert {\bf k} \vert$), we have solved this equation in the forward scattering limit corresponding to not-too-large angular deviations of the wave-vector with respect to its initial direction of emission. The resulting Fokker-Planck equation is a generalization of Fermi's pencil beam equation which accounts for the phenomenon of parallel slowing-down, in addition to angular spread, as a result of elastic scattering of the photons. The perpendicular and parallel spatial parts of this equation were solved separately to obtain the joint evolution of the photon distribution in both space and wave-momentum space. The expression for the angular broadening coefficient $D_{\psi}$, related to the spatial extent of the scattered image of the radio source, is readily recovered from the solution.

We also considered the diffusive regime of spatial transport characterized by the diffusivity $\kappa$ of photons. In both the small-angle and diffusive regimes, we obtained the form of the pulse time profile that results from propagation in a homogeneous scattering shell of length $L$ from the point source location toward the observer; in both cases the pulse shape is characterized by a relatively rapid rise and a slower exponential decay with a characteristic timescale that decreases with frequency. It has been suggested that the temporal decay of the emitted radiation could be the direct consequence of the decay of the source emission through collisional damping of Langmuir waves, hence offering a method to study the temperature of the solar corona based on the observation of type III bursts. The validity of this hypothesis was gauged long ago \citep{1973SoPh...29..197A,1974SoPh...35..153R} and the main criticism remains that the collisional damping time of Langmuir waves does not agree well with the observed decay time; we refer to the discussion of Figure~4 in \cite{2014A&A...572A.111R} for more details on this issue.

While our analytical results account well for the observed pulse shapes of type III bursts, with a fast rise and an exponential decay, the predicted frequency dependence of the decay time $\tau_d$ (Equation~(\ref{tau-d-omega})), in either the weak or strong scattering regimes, does not agree well with observations, which instead show a behavior

\begin{equation}\label{f-scaling}
\tau_{d}(f) \propto f^{-1}
\end{equation}
that is observed to hold over many decades in frequency (see Figure~2 in \cite{1973SoPh...29..197A}, which is based on a collection of measurements from different instruments in the range $0.1-100$~MHz). \cite{1985A&A...150..205S} has discussed a similar discrepancy related to the variation of source size with frequency (which also has an $f^{-1}$ dependence).  It should be noted, however, that in the analysis of the Fokker-Planck equation (\ref{fp-equation-f}), we have set the large-scale refractive term to zero, and hence we did not take into account the role of the large-scale density inhomogeneity. In obtaining analytical solutions to the Fokker-Planck equation (\ref{fp-equation-f}), the wavenumber was taken as a constant of motion and we did not considered the effect of the change in $\vert {\bf k} \vert$ resulting from refraction. In other words, while both the strong ($\overline{\mu}_{r}\ll 1$) and weak ($\overline{\mu}_{r}\sim 1$) scattering regimes are considered in a unified way, our working hypothesis of a spatially uniform index of refraction in the scattering region does not allow for a transition between these two scattering regimes during photon propagation.

We therefore believe that neglect of refraction is the most likely the reason for the discrepancy between the predicted and observed frequency dependencies of $\tau_d$. Indeed, \cite{2018ApJ...857...82K} have used the Monte-Carlo technique developed by \cite{2007ApJ...671..894T} to simulate the arrival time profiles of radio bursts at 1AU. Their numerical results fit rather well STEREO observations in the range $100-1000$ kHz, a low-frequency range in which collisional damping of Langmuir waves and collisional absorption of radio waves are both negligibly small. \cite{2018ApJ...857...82K} show that the fundamental component of the emission decays exponentially, with a frequency-dependent decay time $\tau_d(f)$ that is broadly consistent with Equation~(\ref{f-scaling}). Since none of the fundamental physical processes intrinsic to the source of emission (see Section~\ref{emission-mechanisms}) are included in the point-source transport model of \cite{2018ApJ...857...82K}, the major physical processes controlling the behavior of the burst time profile at these low frequencies must be either scattering or refraction of the emitted radiation.  It is reasonable to assume that these processes also play a similar role at higher frequencies.

Refraction alone can produce a dispersion of the photon arrival times at 1AU: photons emitted in different directions follow different deterministic paths with a varying group velocity and are therefore received at different times. However, the uncorrelated ray paths associated with refraction in a spherically symmetric solar atmosphere \citep{2007ApJ...671..894T} are not expected to yield a well-defined pulse shape such as the one shown in Figure~\ref{numerical-solution}. Indeed, histograms of the arrival times of photons at 1 AU, with and without the statistical correlation effect of scattering included, have been computed by \cite{2007ApJ...671..894T} and compared. The results, presented in their Figure~3, show that a characteristic pulse shape similar to that observed can only emerge under the effect of turbulent scattering, i.e., by randomization of the photon trajectories and hence of the arrival times, the effect studied in Section~\ref{pulse-profile-analysis}.

Combining all the above, then, it is apparent that \emph{both} scattering and refraction are important elements in determining the arrival time profiles of Type~III bursts and their frequency dependence. A complete explanation of how their combination results in the observed scaling law~(\ref{f-scaling}) is therefore an important area for future research.

Finally we have shown the importance of the temporal dependence of plasma density fluctuations close to the source of emission. Photons emitted close to the plasma frequency have a group velocity that is sufficiently low that even slow time variations in the density fluctuations can cause diffusive photon acceleration (a change in frequency) through a Fermi process while also diffusing in space.

Overall, this leads to the following qualitative picture: in the vicinity of the source, scattering of photons causes them to evolve according to a double diffusion in both space and energy (Equation~(\ref{double-diffusion})). They are strongly scattered in angle and hence undergo a spatial diffusion while also being stochastically accelerated by the time-dependent part of the plasma fluctuations. The change in frequency resulting from the stochastic acceleration increases the group velocity from an initial value close to zero for the fundamental mode and hence allows photons to escape diffusively from the emission/acceleration region. Further away from the source, quasi-static variations in the ambient density result in predominantly angular scattering at constant energy (Equation~(\ref{scattering-operator})). All these effects, acting while the photons are transiting from a diffusive regime of acceleration and transport to a regime of relatively free propagation, are important to explain the subsecond variation of the sources observed by LOFAR \citep{2017NatCo...8.1515K}. Detailed analysis of the spatial and temporal properties of radio bursts resulting from photon acceleration and transport is expected to provide a very useful diagnostic of the plasma turbulence intervening inside a few solar radii from coronal sources.

\acknowledgements

NB and AGE were supported by grant NNX17AI16G from NASA's Heliophysics Supporting Research program. EPK was supported by STFC consolidated grant ST/P000533/1.

\bibliographystyle{aasjournal}

\bibliography{bian_et_al_wavescat}

\end{document}